\documentclass[a4paper,10pt]{article}
\usepackage{amsmath,amssymb,amscd,amsthm}
\usepackage[all]{xy}
\begin{document}
\title{ON GHOST FERMIONS}

\author{G. Grensing\\ Institut f\H ur Theoretische Physik \\und Astrophysik \\ 
Universit\H at Kiel \\D-24118 Kiel, 
Germany}

\date{May 2001}

\maketitle

\begin{abstract}

The path integral for ghost fermions, which is heuristically made use of in the Batalin-
Fradkin-Vilkovisky approach to quantization of constrained systems, is derived from first 
principles. The derivation turns out to be rather different from that of physical 
fermions since the definition of Dirac states for ghost fermions is subtle. With these 
results at hand, it is then shown that the nonminimal extension of the 
Becchi-Rouet-Stora-Tyutin operator must be chosen differently from the notorious choice 
made in the literature in order to avoid the boundary terms that have always plagued 
earlier treatments. Furthermore it is pointed out that the elimination of states with 
nonzero ghost number requires the introduction of a thermodynamic potential for ghosts; 
the reason is that Schwarz's Lefschetz formula for the partition function of the time-
evolution operator is not capable, despite claims to the contrary, to get rid of nonzero 
ghost number states on its own. Finally, we comment on the problems of global topological 
nature that one faces in the attempt to obtain the solutions of the Dirac condition for physical states
in a configuration space of nontrivial geometry; such complications give rise to anomalies 
that do not obey the Wess-Zumino consistency conditions.
\end{abstract}
\thispagestyle{empty}
\newpage
\setcounter{page}{1}
\numberwithin{equation}{section}
\section*{Introduction and Summary}

Systems with first class constraints, of which abelian and nonabelian gauge theories are 
prime examples, are rather perfectly understood classically through Marsden-Weinstein 
reduction \cite{Wood 92}. The quantization of such systems is achieved by means of the 
Batalin-Fradkin-Vilkovisky (BFV) approach \cite{Frad 75,Bata 77,Frad 78}, being based on 
the Becchi-Rouet-Stora-Tyutin (BRST) construction which in turn follows from Faddeev's 
formula \cite {Fadd 69} as the essential ingredient; hence, just the opposite strategy is 
pursued since, instead of restricting the phase space, it is enlarged by introducing additional 
ghost degrees of freedom.

But there are still some open problems, such as the construction of physical states with 
finite norm in the operator approach (see, e.g., \cite{Marn 91,Bata 95}). What is also 
missing is a proper understanding of the path integral for ghost fermions, which is only 
formally written down in the BFV approach, without clarifying its origin. Furthermore, it 
is not known in which way the partition function of the BFV system by itself manages  
that the cohomology collapses at zero ghost number \cite{Roge 97}. It is the purpose of 
the present paper to contribute to a solution of these problems.

In particular, we attempt to give a derivation of the path integral for ghost fermions at 
a comparable level of rigor as that invested for the other ones (see, e.g., \cite{Schu 
81}). Let us recall, there are three different types of path integrals which  were 
investigated in the past and are rather well understood by now. These are
\begin{itemize}
\item the Feynman path integral in its original lagrangian version; its hamiltonian 
version \cite{Tobo 56,Davi 63,Garr 66} is, to cite Henneaux (\cite{Henn 85}, p.65), "full 
of subtleties", but it is actually the one being required in the context of constrained 
systems.
\item the coherent state path integral, which in many aspects is simpler than the Feynman 
type of path integral; it has the virtue to admit a rather `coherent' treatment of bosons 
and Dirac fermions \cite{Bere 66,Fadd 76}, but is not applicable to ghost fermions.
\item the Berezin path integral \cite{Bere 71}, which uses the symplectic structure of 
the phase space for fermions and bosons through the Weyl approach to quantization 
\cite{Bere 77,Mari 80}.
\end{itemize}
What has been brought forward \cite{Henn 92} as a possible candidate for a path integral 
of ghost fermions, is Berezin's variant. This, however, leads to boundary conditions, 
being essentially different from that obtained for a Feynman path integral in hamiltonian 
form; but it is the latter type of boundary conditions that is needed here. Hence, 
standard techniques fail for the case at hand. 

We approach the problem by extending earlier work of Marnelius \cite{Marn 87}. For this, 
we first give a construction of the Dirac basis for ghost fermions, which is not at all 
straightforward since it requires the introduction of an unconventional kind of (real) 
coherent states. Then we are able to derive the corresponding path integral by following 
rather standard lines so that the trace and supertrace of a zero ghost number operator 
can be expressed as a functional integral. These matters form the content of the first 
section, which are applied in the second section to the BFV system. There it is shown 
that the nonminimal extension of the BRST operator requires modification in order to 
achieve that the BFV partition function really is invariant under BRST transformation, 
whereas the notorious choice made in the literature \cite{Frad 75,Henn 92} suffers from 
boundary terms \cite{Henn 85,Henn 92,Henn 92a} that destroy its invariance. This 
modification also admits to give a proof of the Fradkin-Vilkovisky \cite{Frad 75} theorem that 
rectifies some weak points of the original version \cite{Bata 77}. We then turn to the 
operator treatment of the BRST approach, where it is known from Schwarz's work \cite{Schw 
89} that it is the supertrace in the ghost sector, which must be used in order to achieve  
that only the contributions from the cohomology groups survive in the partition function.  
Using the results of the first section, we can then express the Lefschetz formula as a 
functional integral which, however, involves the contributions of all cohomologies and 
not, as one wants, of the zero cohomology only. This problem has also been seen and dealt 
with recently in \cite{Roge 97}, but our answer is different. As we believe, the problem 
can only be settled by introducing a thermodynamic potential for the ghosts. Then one can 
get rid of the nonzero cohomologies by isolating that part of the total partition 
function, which is independent of the thermodynamic potential. In the concluding section 
it is demonstrated on the example of abelian Chern-Simons theory in the plane and on the 
torus that, through the integrated version of the Dirac condition for physical states, an 
anomaly \cite{Trei 85} is encountered since the associated group two-cocycle is generally 
non trivial.

\section{GHOST FERMIONS}

We want to model the fermionic analogue of bosonic momentum operators $\hat{p}_i$ and 
generalized coordinate operators $\hat{q}^j$, denoted by $\hat{\zeta}_a$ and 
$\hat{\eta}^b$ in the following. The treatment of these matters in the literature, if 
given at all, is both controversial and incomplete. For example, Berezin and Marinov 
\cite{Bere 77} remark that "in the Grassmann phase space one cannot use the 
coordinate-momentum language, and it is impossible to define analogue of the Feynman 
path integral in the coordinate (or momentum) space." This statement is indeed true for 
physical real 
fermions. As we will show, however, the above verdict may be overcome for ghost fermions. 

What will turn out to be a nontrivial affair is to construct a Dirac basis for such 
unphysical fermions. Recall in this context that a proper definition of (bosonic) Dirac 
kets, which are needed for the Feynman path integral approach to quantization, is a subtle issue 
that requires the concept of Gel'fand triplets \cite{Gelf 64,Bohm 89}. Hence it should 
come to no surprise that also the construction of Dirac states for ghost fermions will 
involve some subtleties.

\subsection{SCHR\"{O}DINGER REPRESENTATION}

It is natural to assume that the operators, corresponding to the real fermionic momentum  
variables $\zeta_a$ and coordinate variables $\eta^b$, must obey the anticommutation 
relations
\begin{equation}\label{relationsunphysicalfermions}
[\hat{\zeta}_a,\hat{\eta}^b]_+=\delta_a{}^b
\end{equation}
where $a,b\in\{1,\ldots,m\}$. The operators $\hat{\zeta}_a$ and $\hat{\eta}^b$ are 
supposed to be selfadjoint, in a sense to be made precise, because the corresponding 
Grassmann variables are real by assumption. It is for this reason, that the factor 
$\sqrt{-1}$ is missing on the right-hand side of the basic anticommutator. 

A straightforward strategy to find a realization of the algebra of operators with the 
above defining relations is to proceed along the lines of the bosonic case. So we 
introduce fermionic `Schr\" odinger' wave functions (cf. also \cite{Thom 88,Holt 90})
\begin{equation}
\psi(\eta)=\sum_{p=0}^{m}\frac{1}{p!}\eta^{a_1}\cdots\eta^{a_p}\,\psi_{a_1\cdots a_p}
\end{equation}
of the Grassmannian configuration space variables $\eta^a$; they are real in the sense 
$(\eta^a)^{\ast}=\eta^a$, with the $\ast$-involution inverting the order of the factors: 
$(\eta^{a_1}\cdots\eta^{a_p})^{\ast}=\eta^{a_p}\cdots\eta^{a_1}$. The completely 
antisymmetric coefficients $\psi_{a_1\cdots a_p}$ are assumed to take complex values. On 
such wave functions, the operators $\hat{\zeta}_a$ and $\hat{\eta}^b$ are defined to act 
as
\begin{equation}\label{hatetajpsietaequaletajpsieta}
\hat{\eta}^b\psi(\eta)=\eta^b\psi(\eta)\qquad\hat{\zeta}_a\psi(\eta)=\frac{\partial}
{\partial\eta^a}\psi(\eta).
\end{equation}
Here, the derivative must necessarily act from the left in order to reproduce the 
fundamental anticommutator. 

We now turn to the definition of selfadjointness for the above operator realization. So a 
sesquilinear form $\langle\psi|\psi^{\prime}\rangle$ on these Grassmann valued wave 
functions must be introduced; a natural choice is
\begin{equation}\label{sesquilinearformunphysicalfock}
\langle\psi|\psi^{\prime}\rangle=\int\,d^{m}\eta\,\psi(\eta)^{\ast}
\psi^{\prime}(\eta)=\langle\psi|\psi^{\prime}\rangle=\sum_{p}\frac{(-1)^{{p \choose 
2}}}{(m-p)!p!}\,\varepsilon^{a_1\cdots a_pa_{p+1}\cdots a_m}\psi^{\ast}_{a_1\cdots 
a_p}\psi^{\prime}_{a_{p+1}\cdots a_m}.
\end{equation}
With respect to this inner product the operators $\hat{\zeta}_a$ and $\hat{\eta}^b$ are 
selfadjoint; on using the rules of Grassmann 
calculus, the proof is by direct verification. There is also a ghost number operator 
\cite{Kugo 79} available
\begin{equation}
\hat{N}=\frac{1}{2}\left(\hat{\eta}^a\hat{\zeta}_a-\hat{\zeta}_a\hat{\eta}^a\right)
\end{equation}
being constructed such that it is skew adjoint with respect to the inner product; it  
counts the momentum operators as $+1$ and the coordinate operators as $-1$.
On the subspace of functions  
$\psi_p(\eta)=\frac{1}{p!}\eta^{a_1}\cdots\eta^{a_p}\,\psi_{a_1\cdots a_p}$ of na\"\i ve 
Grassmann degree $p$ the operator $\hat{N}$ is diagonal with eigenvalue $p-\frac{m}{2}$, 
which is ${m \choose p}$-fold degenerate.

For the investigation of the properties of this sesquilinear form, the above explicit 
expression \eqref{sesquilinearformunphysicalfock} is not very useful. Instead, it is 
advantageous to turn to a (equivalent) Fock type of representation for the wave function
\begin{equation}\label{schroedingerwavefunctionrealfermionsfock}
\psi(\eta)=\sum_{n_1,\ldots,n_m=0}^{1}(\eta^1)^{n_1}\cdots
(\eta^m)^{n_m}\psi_{n_1\cdots n_m}.
\end{equation}
We then obtain the alternative expression
\begin{equation}\label{innerproductfockrepresentationunphysical}
\langle\psi|\psi^{\prime}\rangle=\sum_{n_1\cdots n_m}\,(-1)^{\sum_{i}n_i(i-1)}
\psi_{n_1\cdots n_m}^{\ast}\psi^{\prime}_{\bar{n}_1\cdots\bar{n}_m}
\end{equation}
where $\bar{n}_i=1-n_i $, from which one infers that the sesquilinear form is 
nondegenerate.

However, this nondegenerate sesquilinear form is generally not hermitian (or symmetric 
for real functions) because
\begin{equation}
\langle\psi|\psi^{\prime}\rangle^{\ast}=(-1)^{{m \choose 2}}
\langle\psi^{\prime}|\psi\rangle.
\end{equation}
We could enforce hermiticity by a simple redefinition of the inner product; but in the 
field theoretic case (as well as for functions with real coefficients) this approach is 
not amenable since this would yield an unwanted accumulation of phase factors. So we must 
take both the indefiniteness and the non-hermiticity at face.

The indefiniteness of the inner product reflects the properties of the corresponding 
Clifford algebra with generating elements $\hat{\xi}_{\alpha}$, given by 
\begin{equation}\label{metricunphysicalrealfermions}
\hat{\xi}_a=\hat{\zeta}_a\qquad\quad\hat{\xi}_{m+a}=\hat{\eta}_a
\end{equation}
where $\alpha=1,\ldots,2m$. The defining relations are 
$\hat{\xi}_{\alpha}\hat{\xi}_{\beta}+\hat{\xi}_{\beta}\hat{\xi}_{\alpha}=g_{\alpha\beta}$, 
with the metric tensor
\begin{equation}
g=\begin{pmatrix}0&1_m\\1_m&0\end{pmatrix}
\end{equation}
indeed being indefinite. 

This property of unphysical real fermions is in marked contrast to the properties of 
complex physical fermions. For these, the fundamental anticommutator is 
\begin{equation}
[\hat{\psi}^{\ast}_A(\mathbf{x}),\hat{\psi}^B(\mathbf{y})]_+=\delta_A{}^B
\delta(\mathbf{x},\mathbf{y}).
\end{equation}
Omitting the $x$-dependence and spinor indices $A,B$ altogether, the corresponding (real)  
Clifford algebra generators are
\begin{equation}\label{cliffordgeneratorsphysferm}
\hat{\xi}_1=\frac{1}{\sqrt{2}}(\hat{\psi}+\hat{\psi}^{\ast})\qquad
\hat{\xi}_2=\frac{i}{\sqrt{2}}(\hat{\psi}-\hat{\psi}^{\ast})
\end{equation}
which yield 
\begin{equation}\label{metricphysicalrealfermions}
g=\begin{pmatrix}1&0\\0&1\end{pmatrix}
\end{equation}
that is, a positive definite metric. For these physical real fermions, however, there is 
no natural splitting of $\xi_1$ and $\xi_2$ into a coordinate and momentum; the choice of 
a real polarization would destroy rotational invariance. Only the holomorphic 
polarization is available, which is made use of in the standard coherent state 
representation \cite{Fadd 76}.

On the other hand, for real unphysical fermions, one could try to pass in analogy to 
\eqref{cliffordgeneratorsphysferm} to operators $a=(\zeta-i\eta)/\sqrt{2}$ and 
$a^{\ast}=(\zeta+i\eta)/\sqrt{2}$ obeying $(a^{\ast})^{\ast}=a$; however, since 
$[a,a^{\ast}]_+=0$ they cannot be interpreted as fermionic creation and annihilation 
operators. Hence, for ghost fermions a complex structure does not make sense.

\subsection{VECTOR SPACE REALIZATION}

We want to give a conventional matrix realization of the operators  $\hat{\zeta}_a$ and 
$\hat{\eta}^b$ on a $2^m$-dimensional complex vector space, i.e. without taking recourse 
to Grassmann variables. This construction will be needed in the following subsection.

For this purpose, we choose a complex linear space of dimension $2^m$ with basis 
$|n_1,\ldots,n_m\rangle$ where $n_a=0,1$, the general element of which we write in the 
form
\begin{equation}\label{verticalbarpsirangleintermsofthevectorspacebasis}
|\psi\rangle=\sum_{n_1\cdots n_m}|n_1,\ldots,n_m\rangle\psi(n_1,\ldots,n_m).
\end{equation}
Taking \eqref{hatetajpsietaequaletajpsieta} as a guiding principle, the action of the 
momentum and coordinate operators on the basis is defined to be
\begin{equation}
\begin{split}\label{matrixrealizationmomentaandcoordinates}
\hat{\zeta}_a|n_1,\ldots,n_a,\ldots,n_m\rangle&=(-1)^{n_1+\cdots+n_{a-
1}}\bar{n}_a|n_1,\ldots,\bar{n}_a,\ldots,n_m\rangle
\\
\hat{\eta}_a|n_1,\ldots,n_a,\ldots,n_m\rangle&=(-1)^{n_1+\cdots+n_{a-
1}}n_a|n_1,\ldots,\bar{n}_a,\ldots,n_N\rangle.
\end{split}
\end{equation}
Let us introduce the special state $|0\rangle=|0,\ldots,0\rangle$, satisfying 
$\hat{\eta}^a|0\rangle=0$ for all $a\in\{1,\ldots,m\}$, and with the help of which we can 
generate the whole basis according to
\begin{equation}\label{definitionbasisfromlowestweightstate}
(\hat{\zeta}_1)^{n_1}\cdots(\hat{\zeta}_m)^{n_m}|0\rangle=|n_1,\ldots,n_m\rangle.
\end{equation}
Of course, the eqs. \eqref{matrixrealizationmomentaandcoordinates} realize nothing else 
but the standard Fock space construction of the creation operators $\hat{\zeta}_a$ and 
the destruction operators $\hat{\eta}^a$; what is crucially different, however, this is 
the inner product on the Fock space. In the present case it must be chosen such that the 
creation and annihilation operators are selfadjoint, whereas in the standard case they 
are adjoint to one another. For this purpose we imitate 
\eqref{innerproductfockrepresentationunphysical} and define the (nonstandard) inner 
product to be
\begin{equation}\label{innerproductnormalization}
\langle n_m,\ldots,n_1|n^{\prime}_1,\ldots,n^{\prime}_m\rangle=\,(-1)^{\sum_{a}n_a(a-1)}
\delta_{\bar{n}_1n^{\prime}_1}\cdots\delta_{\bar{n}_mn^{\prime}_m}.
\end{equation}
Again, this is nondegenerate, but neither hermitian nor positive definite; in particular, 
all basis vectors have norm zero. This is the explicit realization of the vector space 
together with its indefinite inner product, being implicitly encountered in the operator 
approach to the BRST algebra \cite{Kugo 79,Nish 84}. Below we shall have need of the 
particular basis vector 
\begin{equation}\label{barzerorangle}
|\bar{0}\rangle=(-1)^{{m \choose 2}}|1,\ldots,1\rangle\end{equation}
which is of `highest weight' with respect to the destruction operators $\hat{\eta}^a$ and 
normalized such that $\langle\bar{0}|0\rangle=1$ holds. What remains to prove is that the 
matrix representation of the momentum and coordinate operators 
\eqref{matrixrealizationmomentaandcoordinates} is selfadjoint with respect to 
\eqref{innerproductnormalization}; the proof is by direct verification. The last point to 
be discussed is the completeness relation. For this purpose, we use
\begin{equation}
\langle n_m,\ldots,n_1|\psi\rangle=\,(-1)^{\sum_{a}n_a(a-1)}
\psi(\bar{n}_1,\cdots,\bar{n}_m)
\end{equation}
and this in turn gives
\begin{equation}\label{completenessrelationfockbasisrealfermins}
\sum\limits_{n_1\cdots n_m}\,(-1)^{\sum_{a}\bar{n}_a(a-1)}|n_1,\ldots,n_m\rangle
\langle \bar{n}_m,\ldots,\bar{n}_1|=1_{2^m}
\end{equation}
which is the result sought for.

It is quite remarkable that along these lines one can circumvent the general 
representation theory of Clifford algebras on vector spaces of dimension $2m$ with an 
inner product of zero signature. As is known, the general approach (see \cite{Roe 88}) makes essential 
use of the representation theory of finite groups to prove the existence of a (unique) 
representation of dimension $2^m$, which we here have constructed explicitly.

\subsection{DIRAC STATES AND THEIR DUALS}

Ultimately, what we want is a path integral for these unphysical real fermions, which 
must be derived from first principles; but for this it is mandatory to have available a 
Dirac basis. From the treatment of standard (complex) coherent states (cf. also 
\cite{Marn 87}), we are acquainted with the definition
\begin{equation}
|\eta\rangle=\exp(\hat{\zeta}\cdot\eta)|0\rangle
\end{equation}
and the action of the coordinate and momentum operators on this Dirac basis is
\begin{equation}
\hat{\eta}^a|\eta\rangle=\eta^a|\eta\rangle
\qquad\qquad\hat{\zeta}_a|\eta\rangle=\frac{\partial_r}{\partial\eta^a}|\eta\rangle
\end{equation}
where the subscript $r$ ($l$) denotes the right (left) derivative. 

However, now the construction of the matrix realization of the preceding paragraph makes 
no sense, unless we attach a degree to the basis vectors. To see this recall the fact  
that there exists a state of `lowest weight' $|0\rangle$, which is annihilated by the 
coordinate operators, and from which the complete set of basis vectors 
$|n_1,\ldots,n_m\rangle$ can be generated according to \eqref{definitionbasisfromlowestweightstate} by repeated application of the momentum operators. Hence, if we attach 
the degree zero to the state $|0\rangle$, then the assignment of the degree $\sum_an_a$ 
to $|n_1,\ldots,n_m\rangle$ makes the Dirac ket basis vector $|\eta\rangle$ an even 
quantity. Expressed in terms of the now Grassmann valued vector space basis, its 
expansion takes the form
\begin{equation}
|\eta\rangle=\sum_{n_1\cdots n_m}|n_1,\ldots,n_m\rangle(\eta^m)^{n_m}\cdots(\eta^1)^{n_1}
\end{equation}
where here and below the ordering of the factors is essential.

The subtle point is to construct the corresponding bra vector $\langle\eta|$, which we 
want to yield the Grassmann $\delta$-function 
\begin{equation}\label{langleetatimesetaprimerangle}
\langle\eta|\eta^{\prime}\rangle=\delta(\eta-\eta^{\prime})=(-1)^m(\eta-
\eta^{\prime})^1\cdots(\eta-\eta^{\prime})^m
\end{equation}
in analogy with the bosonic case; this has the na\"\i ve degree $m$ so that the bra 
$\langle\eta|$ must have the same na\"\i ve degree. It is for this reason, that we 
cannot choose the conventional adjoint of $|\eta\rangle$; instead, we must define
\begin{equation}
\langle\eta|=\langle\bar{0}|(\hat{\eta}^1\cdots\hat{\eta}^m)\exp(\eta\cdot\hat{\zeta}).
\end{equation}
The point of crucial importance with this definition is that we must give the dual 
$\langle\bar{0}|$ of $|0\rangle$ the degree zero in order to make sense, whereas the 
conventional counting for $|\bar{0}\rangle$ according to \eqref{barzerorangle} yields 
$m$. Hence, we must alter the assignment of a degree to the adjoint basis since otherwise 
the quantity $\langle\bar{0}|0\rangle$ would be Grassmann valued.

This can consistently be done as follows. For this purpose, let us introduce the operator
\begin{equation}
\hat{G}=\sum_a\hat{\zeta}_a\hat{\eta}^a
\end{equation}
which counts what we call the \emph{Grassmann degree}. Its action on the basis (see 
\eqref{definitionbasisfromlowestweightstate}) is 
\begin{equation}
\hat{G}|n_1,\ldots,n_m\rangle=(\sum_an_a)|n_1,\ldots,n_m\rangle.
\end{equation}
which is the same as the conventional na\"\i ve degree. Hence, for states we distinguish 
between the ghost number, being counted by $\hat{N}$, and the ghost degree, being counted 
by $\hat{G}$; its adjoint is $\hat{G}^{\ast}=m-\hat{G}$. Consequently, for its action on 
the adjoint basis  $\langle0|(\hat{\zeta}_m)^{n_m}\cdots(\hat{\zeta}_1)^{n_1}$, the 
conventional degree of which is $\sum_an_a$, we obtain instead 
$$\langle n_m,\ldots,n_1|\hat{G}=\langle n_m,\ldots,n_1|(\sum_a\bar{n}_a).$$
Furthermore, we pass to the dual basis (see \eqref{innerproductnormalization} and 
\eqref{completenessrelationfockbasisrealfermins})
\begin{equation}
\langle\overline{n_m,\ldots,n_1}|=\,(-1)^{\sum_{a}\bar{n}_a(a-
1)}\langle\bar{n}_m,\ldots,\bar{n}_1|=\langle\bar{0}|(\hat{\eta}^m)^{n_m}\cdots
(\hat{\eta}^1)^{n_1}
\end{equation}
with the properties 
\begin{equation}
\langle\overline{n_m,\ldots,n_1}|n^{\prime}_1,\ldots,n^{\prime}_m\rangle=\delta_
{n_1nn^{\prime}_1}\cdots\delta_{n_mn^{\prime}_m}
\end{equation}
\begin{equation}
\sum_{n_1\cdots n_m}|n_1,\ldots,n_m\rangle\langle\overline{n_m,\ldots,n_1}|=1_{2^m}.
\end{equation}
This dual basis then has the Grassmann degree
\begin{equation}
\langle\overline{n_m,\ldots,n_1}|\hat{G}=\langle\overline{n_m,\ldots,n_1}|
(\sum_an_a).
\end{equation}
With this assignment, both $|0\rangle$ and $\langle\bar{0}|$ have degree zero, and thus 
$\langle\eta|$ has Grassmann degree $m$, as we wanted to achieve. Had we assigned  to 
$\langle\bar{0}|$ the conventional degree $m$, then $\langle\bar{0}|0\rangle$ would also 
be of degree $m$, and we could not give this quantity the numerical value one.

It is straightforward now to show that the coordinate and momentum operators act on the 
Dirac bra vectors as
\begin{equation}
\langle\eta|\hat{\eta}^j=\langle\eta|\eta^j\qquad\langle\eta|\hat{\zeta}_i=
-\frac{\partial_r}{\partial\eta^i}\langle\eta|.
\end{equation}
Furthermore, one can prove the conjectured normalization property 
\eqref{langleetatimesetaprimerangle} of the Dirac basis by means of the explicit form
\begin{eqnarray}
\langle\eta|&=&\langle\bar{0}|(\hat{\eta}^1-\eta^1)\cdots(\hat{\eta}^m-\eta^m)\\ 
\nonumber
&=&(-1)^{{m \choose 2}}\sum_{n_1\cdots n_m}(-1)^{\sum_a\bar{n}_a(m-
a+1)}(\eta^1)^{\bar{n}_1}\cdots(\eta^m)^{\bar{n}_m}\langle\overline{n_m,\ldots,n_1}|
\end{eqnarray}
in its first version. The second version is needed to reduce the proof of the 
completeness relation for the Dirac basis, which is 
\begin{equation}
(-1)^m\int\,d^{m}\eta|\eta\rangle\langle\eta|=1
\end{equation}
to the completeness relation \eqref{completenessrelationfockbasisrealfermins} of the Fock 
basis. 

Let us relate these results to the Schr\"odinger wave function approach of the last but 
one subsection. For $\psi(\eta)$ (see \eqref{schroedingerwavefunctionrealfermionsfock}) 
with 
\begin{equation}
|\psi\rangle=\int\,d^{m}\eta\,|\eta\rangle\psi(\eta)
\end{equation}
we obtain from the completeness relation
\begin{equation}
\psi(\eta)=(-1)^m\langle\eta|\psi\rangle
\end{equation}
and for the coefficients $\psi_{n_1\ldots n_m}$, this gives
\begin{equation}
\psi_{n_1\ldots n_m}=(-1)^{{m+1 \choose 2}}(-1)^{\sum_an_a(m-
a+1)}\psi(\bar{n}_1,\ldots,\bar{n}_m)
\end{equation}
which now come equipped with a Grassmann degree.

We end the discussion of the Dirac basis over configuration space with an investigation of 
the trace of an operator $\hat{O}$ with zero ghost number; this we define by means of the 
dual basis to be
\begin{equation}
\mathrm{Tr}\,\hat{O}\,=\,\sum_{n_1\cdots 
n_m}\langle\overline{n_m,\ldots,n_1}|\hat{O}|n_1,\ldots,n_m\rangle.
\end{equation}
One can also introduce a supertrace, defined by
\begin{equation}
\mathrm{Str}\,\hat{O}\,=\,\sum_{n_1\cdots n_m}(-
1)^{\sum_an_a}\langle\overline{n_m,\ldots,n_1}|\hat{O}|n_1,\ldots,n_m\rangle.
\end{equation}
By means of the Dirac basis, these traces can be expressed in the form
\begin{equation}\label{traceandsupetraceghostfermionsincoordinatespace}
\mathrm{Tr}\,\hat{O}\,=\,\int\,d^{m}\eta\,\langle-
\eta|\hat{O}|\eta\rangle\qquad\qquad\quad
\mathrm{Str}\,\hat{O}\,=\,\int\,d^{m}\eta\,\langle\eta|\hat{O}|\eta\rangle
\end{equation}
as follows by a straightforward computation.

One can as well construct a Dirac basis in momentum space and introduce Fourier 
transformation; the relevant formulae are collected in an appendix.

\subsection{FEYNMAN TYPE PATH INTEGRAL}

Having available the Dirac basis, the path integral treatment of the time-evolution 
operator for these unphysical fermions is rather straightforward; it closely follows the 
analogous bosonic case \cite{Fadd 76}, and so we may be brief.

We assume the Hamiltonian $\hat{H}=H(\hat{\zeta},\hat{\eta})$ to be an even operator, the 
ordering being prescribed such that the momentum operators are placed to the left of the 
coordinate operators. The transition amplitude
\begin{equation}
\langle\eta^{\prime\prime}|\exp-i\hat{H}(t^{\prime\prime}-
t^{\prime})|\eta^{\prime}\rangle=\langle t^{\prime\prime},\eta^{\prime\prime}|\eta^
{\prime},t^{\prime}\rangle
\end{equation}
can be written in the form of a path integral as follows
\begin{equation}
\langle t^{\prime\prime},\eta^{\prime\prime}|\eta^
{\prime},t^{\prime}\rangle=
\end{equation}
$$\lim_{\varepsilon \rightarrow 0}\int d\zeta_{N+1}\cdot d\zeta_Nd\eta_N\cdots 
d\zeta_1d\eta_1\exp i\sum\limits_{n=0}^{N}\left(i\zeta_{n+1}\cdot(\eta_{n+1}-\eta_n)-
\varepsilon H(\zeta_{n+1},\eta_n)\right)$$
where $\eta_0=\eta^{\prime}$ and $\eta_{N+1}=\eta^{\prime\prime}$; note that there is an 
excess of one momentum integration. In formal continuum notation, this reads as
\begin{equation}\label{actionunphysicalrealfermions}
\langle t^{\prime\prime},\eta^{\prime\prime}|\eta^
{\prime},t^{\prime}\rangle=\int\limits_{\eta^{\prime}}^{\eta^{\prime\prime}}
D[\zeta,\eta]\exp i\int_{t^{\prime}}^{t^{\prime\prime}}dt\left(i\zeta\cdot\dot{\eta}-
H(\zeta,\eta)\right).
\end{equation}
As to be expected, the result looks rather similar to the bosonic Feynman path integral 
in hamiltonian form; we stress that only the discrete version, with the limit 
$\varepsilon \rightarrow 0$ taken afterwards, is well defined.

As an application, the transition amplitude can be computed exactly for a selfadjoint 
Hamiltonian of the form
\begin{equation}
\hat{H}(t)=\sqrt{-1}\hat{\zeta}_a\omega^a{}_b(t)\hat{\eta}^b
\end{equation}
with $\omega(t)$ a real square $m$-matrix, which is assumed to be symmetric and may 
depend explicitly on time. Performing in the discrete version the integrations over $(\zeta_n,\eta_n)$ successively for $n=1,\ldots,N$, one ends up with
\begin{eqnarray}\nonumber
\langle\eta^{\prime\prime}|P\exp-
i\int_{t^{\prime}}^{t^{\prime\prime}}dt\hat{H}(t)|\eta^{\prime}\rangle&=&
\lim_{\varepsilon \rightarrow 0}\int d\zeta_{N+1}\exp\left(-
\zeta_{N+1}\cdot\eta_{N+1}+\zeta_{N+1}\cdot e^{\varepsilon\omega_N}\cdots 
e^{\varepsilon\omega_0}
\eta_0\right)\\
&=&\int\,d\zeta^{\prime\prime}\,\exp\left(-
\zeta^{\prime\prime}\cdot\eta^{\prime\prime}+\zeta^{\prime\prime}\cdot 
Pe^{\int_{t^{\prime}}^{t^{\prime\prime}}\omega(t)dt}\eta^{\prime}\right)
\end{eqnarray}
where $\omega_n=\omega(t_n)$ and $P$ signifies the time ordering; the remaining integration over 
$\zeta^{\prime\prime}=\zeta_{N+1}$ can also be done, and we obtain
\begin{equation}
\langle t^{\prime\prime},\eta^{\prime\prime}|\eta^{\prime}
,t^{\prime}\rangle=\delta(\eta^{\prime\prime}-
Pe^{\int_{t^{\prime}}^{t^{\prime\prime}}\omega(t)dt}\eta^{\prime}).
\end{equation}
Hence, only the `classical' solution contributes to the transition amplitude. Finally, we 
can compute, e.g., the supertrace and obtain 
\begin{equation}
\mathrm{Str}\,P e^{-i\int_{t^{\prime}}^{t^{\prime\prime}}\hat{H}(t)dt}=\left|1-
Pe^{\int_{t^{\prime}}^{t^{\prime\prime}}\omega(t)dt}\right|
\end{equation}
where here and below $|A|$ denotes the determinant of a square matrix $A$.

\section{PATH INTEGRAL QUANTIZATION OF CONSTRAINED SYSTEMS}

The path integral for ghost fermions is heuristically made use of in the BVF 
quantization of systems with first order constraints \cite{Frad 75,Henn 92}, without 
specifying its properties. We will show that it is the supertrace, which is used in this 
context, and what the reasons are why this must be so. Furthermore, we comment on the 
proof of the 
Fradkin-Vilkovisky theorem \cite{Bata 77}, which can be improved so as to stand 
objections. 

So let a finite-dimensional Hamiltonian system be given, being subject to $m$ bosonic 
first class constraints $\varphi_a(p,q)$; these are chosen to be momentum maps of an $m$-
dimensional Lie algebra \cite{Wood 92} so that 
$\{\varphi_a,\varphi_b\}=C^c{}_{ab}\varphi_c$ holds where the $C^c{}_{ab}$ are the 
structure constants; the Hamiltonian $H(p,q)$ is assumed to commute with the constraints. 
Furthermore, the auxiliary constraints $\chi^a$ are chosen to be holonomic since the 
constraints are cotangent lifts that are linear homogenous in the momenta; consistency 
then dictates the determinant $|\{\chi,\varphi\}|$ to be nonvanishing. An element of the 
extended supersymmetric phase space is denoted by 
$\xi=(p,q,\mu,\lambda,i\zeta,\eta,i\zeta^{\times},\eta^{\times})$, where the $\lambda^a$ 
denote the Lagrange multipliers with corresponding momenta $\mu_a$; the $\eta^a$ and 
$\zeta_a$ the ghost coordinates and momenta, and analogously $\eta^{\times a}$ and 
$\zeta^{\times}{\!}_b$ for the antighosts. According to Fradkin and Vilkovisky \cite{Frad 
75}, the path integral for this system is
\begin{equation}\label{functionalintegralfv}
Z_{\chi}=\int_{\mathrm{PBC}}\,d[p,q]d[\mu,\lambda]d[\zeta,\eta]
d[\zeta^{\times},\eta^{\times}]
\exp iS_{\chi}
\end{equation}
where the extended action takes the form
\begin{equation}\label{actionfv}
S_{\chi}\,=\,\int_{t^{\prime}}^{t^{\prime\prime}}\,dt\left(p_i\dot{q}^i+\mu_a
\dot{\lambda}^a+i\zeta_a\dot{\eta}^a+i\zeta^
{\times}{\!}_a\dot{\eta}^{\times a}\,-H\,-i\{\Omega,\phi\}\right).
\end{equation}
The meaning of the subscript PBC on the functional integration will be explained later; 
also the reasons for the rather special notation will become apparent below. 
\footnote{The relation to the notation of Henneaux and Teitelboim \cite{Henn 92} is 
$G_a\equiv\varphi_a$ for the constraints, $\lambda_a\equiv\lambda_a$  and 
$b_a\equiv\mu_a$ for the multipliers and their conjugate momenta, $\eta_a\equiv\eta_a$ 
and $\mathcal{P}_a\equiv\zeta_a$ for the ghosts, and 
$\overline{C}_a\equiv\eta^{\times}{}_a$ and $\rho_a\equiv\zeta^{\times}{}_a$ for the 
antighosts.} 
What remains to be specified is the BRST generator $\Omega$ and the gauge-fixing fermion 
$\phi$. Here we depart from the standard choice since we take $\Omega$ to be
\begin{equation}\label{brsqcorrected}
\Omega\,=\,\varphi_a(p,q)\eta^a+\frac{i}{2}\zeta_aC^a{}_{bc}\eta^b\eta^c+\mu_a
\eta^{\times a}
\end{equation}
whereas in the literature the nonminimal term is $\zeta^{\times a}\mu_a$; we shall 
comment on this discrepancy in a moment. Consequently, our gauge-fixing fermion differs  
as well from the standard choice:
\begin{equation}\label{gaugefixingfermionfinalversion}
\phi=\zeta_a\lambda^a+\zeta^{\times}{\!}_a(\chi^a-\frac{\xi}{2}\mu^a).
\end{equation}
Let us first note that, on integrating out the antighost coordinates $\eta^{\times}$ 
and the ghost momenta $\zeta$, the conventional form of the partition function in the 
derivative gauge is obtained; in particular, for the Yang-Mills case, the Faddeev-Popov 
path integral \cite{Fadd 67,Fadd 80} is regained. Hence, our modification does not alter 
the final results.

The essential differences come in if the basic properties of the path integral are 
investigated. The main point in the original work of Fradkin and Vilkovisky \cite{Frad 
75} is the statement that the functional integral for a system with first class 
constraints, as shown in \eqref{functionalintegralfv} and \eqref{actionfv}, does not 
depend on the special choice of the gauge-fixing fermion. A proof of this statement, 
known as the Fradkin-Vilkovisky theorem, was later given by Batalin and Vilkovisky 
\cite{Bata 77}, and since then it has often been repeated. 

The actual proof, however, requires modification for two reasons. The first is that the 
notorious choice made for the nonminimal contribution to the BRST 
charge in the literature is the term $\zeta^{\times a}\mu_a$, which is quadratic in the momenta; it is to 
be contrasted with our choice \eqref{brsqcorrected}, being linear in the momenta. This is 
a crucial point, since a term quadratic in the momenta prevents the functional integral 
from being invariant under BRST transformations. The reason is that nonvanishing boundary 
terms get involved, which destroy its invariance. These boundary terms were discussed by 
Henneaux and Teitelboim (\cite{Henn 85,Henn 92,Henn 92a}, see also \cite{Fadd 69}) in the 
attempt to invent boundary conditions which enforce the vanishing of the terms in 
question. For our choice of the nonminimal term, such boundary terms will be absent. 
Second, for the transformation introduced by Batalin and Vilkovisky in the attempt to 
demonstrate the independence of the choice of the gauge fermion there is no reason to 
believe the extended action to be invariant since the corresponding generating function 
yields a transformation, which is nonlocal. Below we give a proof of the 
Fradkin-Vilkovisky theorem which does not suffer from these defects.

Let us first comment on the boundary term; for this purpose,  
the following simplifying notation is introduced (see, e.g., \cite{Dewi 84}). We 
collectively denote coordinates of  superconfiguration space by 
$z=(q,\lambda,\eta,\eta^{\times})$ and the corresponding momenta by 
$\pi=(p,\mu,i\zeta,i\zeta^{\times})$ so that the BFV partition function takes the form
\begin{equation}
Z_{\phi}(t_2,t_1)\,=\,\int\,d[\pi,z]\,\exp i\int_{t_1}^{t_2}dt(\pi\dot{z}-
H_{\phi}(\pi,z))
\end{equation}
with
\begin{equation}
H_{\phi}=H+i\{\Omega,\phi\}. 
\end{equation}
Here the BRST charge $\Omega=\Omega^{\ast}$ is an odd phase space function with 
$\{\Omega,\Omega\}=0$ which commutes with the Hamiltonian $H$ in the sense 
$\{\Omega,H\}=0$, and as such is a conserved quantity; it contains the original first 
class constraints $\varphi_a$, whereas the gauge-fixing fermion $\phi$ depends on the 
gauge conditions $\chi^a$. Under a supercanonical transformation with (even) 
infinitesimal generating function $\delta G$, the coordinates and momenta transform 
according to 
\begin{equation}
z^A\mapsto z^{A}+\frac{\partial_l\delta G }{\partial\pi_A}\qquad\qquad \pi_A\mapsto\pi_A-
\frac{\partial_r\delta G}{\partial z^A}
\end{equation}
and the partition function is transformed into
\begin{equation}
Z_{\phi}(t_2,t_1)\,=\,\int\,d[\pi,z]\,\exp i\int_{t_1}^{t_2}dt\left(\pi\dot{z}-
H_{\phi}(\pi,z)+\frac{d}{dt}(\pi_A\frac{\partial_l\delta G}{\partial\pi_A}-\delta G)-
\{\delta G,H_{\phi}\}\right)
\end{equation}
where we have used that the super--Liouville measure $d(\pi,z)$ is invariant against general supercanonical BRST transformations. Hence, for the functional integral to be invariant, we must also guarantee that the extended Hamiltonian is invariant, i.e. $\{\delta G,H_{\phi}\}=0$, and that the boundary term 
vanishes
\begin{equation}
\left(\pi_A\frac{\partial_l\delta G}{\partial\pi_A}-\delta G\right)\Big|^{t_2}_{t_1}=0.
\end{equation}
In the present case, the supercanonical transformations have the special form $\delta 
G=\delta\theta \Omega$ with $\Omega$ the nilpotent BRST generator and $\delta\theta$ a 
purely imaginary Grassmann parameter. Then the first requirement is met by construction. 
As to the boundary term, this vanishes if $\delta G$ is linear 
homogeneous in the supermomenta; correspondingly, boundary terms are absent. It is this reason where our request for the BRST generator to be linear in the supermomenta comes from.

Actually, one could weaken the above requirement. A linear, but inhomogeneous generating 
function of the form $G=X^A(z)\pi_A + \Lambda(z)$ would also do since $z^A(t_1)=z^A(t_1)$ on account of the fact that the functional integral is the trace of the time evolution operator in the bosonic and 
the supertrace in the fermionic sector. In particular, for a Yang-Mills theory, such an inhomogeneous term is absent; if it were present, one would face the possibility of an anomalous symmetry.

For the proof of the Fradkin-Vilkovisky theorem, it will turn out to be advantageous to 
use symplectic notation. An element of the supersymmetric phase space is denoted by 
$\xi=(\pi,z)$ and the graded Poisson bracket is
\begin{equation}
\{f,g\}=\frac{\partial_rf}{\partial\xi^{\alpha}}\omega^{\alpha\beta}\frac{\partial_lg}
{\partial\xi^{\beta}}
\end{equation}
where the (co-)symplectic supermatrix with entries $\omega^{\alpha\beta}$ is 
antisupersymmetric. Consider then the transformation 
$\xi^{\prime\alpha}=\xi^{\alpha}+\delta\theta(\xi)\{\Omega(\xi),\xi^{\alpha}\}$
where the Grassmann parameter $\delta\theta$ is now taken to be $\xi$-dependent. 
Accordingly, this is not a canonical transformation, as can be seen by computing the 
superdeterminant of the superjacobian, which is
\begin{equation}
\mathrm{Sdet}\left(\frac{\partial_r\xi^{\prime}}{\partial\xi}\right)=1-
\{\delta\theta,\Omega\}=\exp-\{\Omega,\delta\theta\}.
\end{equation}
We now choose
\begin{equation}\label{choicedeltathetalocal}
\delta\theta(\xi)=i\delta\phi(\xi)\Delta t
\end{equation}
which indeed is a purely imaginary quantity, and so we find for the functional measure in 
its defining discrete version
\begin{equation}
\prod_{n=0}^N\mathrm{Sdet}\left(\frac{\partial_r\xi^{\prime}}{\partial\xi}\right)_n
=\exp-i\sum_n\Delta t_n\{\Omega,\delta\phi\}(\xi_n)
\equiv\exp-i\int_{t_1}^{t_2}dt\{\Omega,\delta\phi\}\quad:\,\Delta t\rightarrow 0
\end{equation}
Let us comment at this point on the notorious choice made in the literature, which instead 
of \eqref{choicedeltathetalocal} is  
$$\delta\theta[\xi]=i\int_{t^{\prime}}^{t^{\prime\prime}}\delta\phi(\xi(t))dt.$$ 
With this construct, however, $\delta G$ is a 
nonlocal generating function that makes no sense in the Hamiltonian formalism. What 
remains finally to be shown is that the transformation
$$\xi^{\prime\alpha}_n=\xi^{\alpha}_n-
i\delta\phi(\xi_n)\{\Omega(\xi_n),\xi^{\alpha}_n\}\Delta t_n$$
leaves the discrete action 
$$\sum_{n=0}^N\left(\pi_n(z_{n+1}-z_n)-\Delta t_nH(\pi_n,z_{n-1})-i\Delta 
t_n\{\Omega,\phi\}(\pi_n,z_{n-1})\right)$$
invariant; but there is nothing to prove since $\delta\pi_n$ and $\delta z_n$ are 
proportional to $\Delta t_n$, and so yield no contribution in the continuum limit. Hence, 
again there are no boundary terms to be discussed away. We thus have proven that 
$Z_{\phi}(t_2,t_1)=Z_{\phi+\delta\phi}(t_2,t_1)$ 
and since finite changes are obtained by exponentiation, the partition function is 
independent of the choice of the gauge-fixing fermion. But this statement requires  
qualification; it only holds for homotopic gauge fermions.
 
\section{OPERATOR APPROACH TO BFV SYSTEMS}

Up to now we have not commented on the paths that enter the BVF path integral. In order 
to discuss this question, we need the operator treatment to BRST quantization. Hence, the 
quantity of interest is the time-evolution operator
\begin{equation}
\hat{U}_{\chi}(t)=\exp -i(\hat{H}+[\hat{\Omega},\hat{\phi}])t
\end{equation}
which is BRST invariant. We assume the Hamiltonian $\hat{H}$ to be Weyl ordered so that 
the midpoint rule is the correct prescription in the path integral. The gauge-fixing 
fermion presents no ordering ambiguities, but the BRST operator does. These are also 
overcome by means of the Weyl ordering, which results in the symmetric ordering for the 
constraints $\varphi_a(p,q)$ since these are linear homogeneous in the momenta by 
assumption; then the algebraic properties of the constraints remain unaltered at the 
operator level. Furthermore, if we define the ghost number operator to be 
\begin{equation}
\hat{N}=\frac{1}{2}(\hat{\eta}^a\hat{\zeta}_a-\hat{\zeta}_a\hat{\eta}^a)+
\frac{1}{2}(\hat{\eta}^{\times a}\hat{\zeta}^{\times}{\!}_a-
\hat{\zeta}^{\times}{\!}_a\hat{\eta}^{\times a})
\end{equation}
then $\hat{\Omega}$ has ghost number $+1$ and $\hat{\phi}$ ghost number $-1$ so that the 
time-evolution operator has ghost number zero. 

For such operators, we know from Schwarz's work \cite{Schw 89} that the supertrace over 
the extended state space reduces to the supertrace over the cohomology groups. In explicit 
terms, let $\hat{O}$ be a BRST invariant operator of zero ghost number. Since the number 
operator defines a grading of the (extended) state space $V=\oplus_l\,V_l$ with $-m/2\leq 
l\leq+m/2$, we can define the restriction $\hat{\Omega}_l$ of the BRST operator to the 
subspace $V_l$; then the $l$th cohomology group is defined as the quotient 
$H^l(\Omega)=\mathrm{Ker}(\hat{\Omega}_l)/\mathrm{Im}(\hat{\Omega}_{l-1})$. This is the 
subspace of $V_l$, consisting of physical states $\Omega \psi=0$ modulo exact states. The 
Lefschetz formula \cite{Schw 89} then says that
\begin{equation}
\mathrm{Str}\hat{O}=\underset{l}{\oplus}(-
1)^l\mathrm{Tr}_{V_l}\hat{O}=\underset{l}{\oplus}(-1)^l\mathrm{Tr}_{H^l(\Omega)}\hat{O}
\end{equation}
since the contributions from non-closed states cancel against those from exact 
states.

Hence, in the ghost fermion sector we must choose the supertrace of the time-evolution 
operator
\begin{equation}
Z_{\chi}(t)=\mathrm{Tr}_{\mathrm{B}}\mathrm{Str}_{\mathrm{GF}}\exp -
i(\hat{H}+[\hat{\Omega},\hat{\phi}])t
\end{equation}
since we want only those states to contribute that obey the generalized Dirac constraint 
$\hat{\Omega}\psi(q,\lambda,\eta,\eta^{\times})=0$. It is of crucial importance to note 
that the traces are taken over the full extended state space, since the restriction to 
the cohomological subspaces is automatic, so that no normalization problems are 
encountered for the partition function. Furthermore, as we have shown in the last but one 
section (see \eqref{traceandsupetraceghostfermionsincoordinatespace}), it is the 
supertrace for the ghost fermions that translates into periodic boundary conditions in 
the functional integral, and this is the meaning of the subscript PBC in 
\eqref{functionalintegralfv}. Also note that the boundary values 
$\eta(t^{\prime\prime})=\eta(t^{\prime})$ and 
$\eta^{\times}(t^{\prime\prime})=\eta^{\times}(t^{\prime})$ are integrated over and thus 
cannot be chosen to be zero, as is done elsewhere.

In addition to the physical state condition, a further restriction is needed in order to 
eliminate negative norm states. This is usually achieved by requiring physical states to 
have ghost number zero. But often this requirement is too restrictive; e.g., for the open 
bosonic string (see \cite{Gree 88}) the relevant cohomologies are at the values $\pm1/2$. 
In general, the correct choice of the relevant cohomology group depends on the system 
under consideration (see also the concluding remark in \cite{Kala 91}); there is no model 
independent proof of a no-ghost theorem.

We return to the BFV system of the form considered in the preceding section. In this 
case, the relevant cohomology group is indeed given by $H^0(Q)$ since the total number of 
constraints is $2m$, i.e., an even number. Nevertheless, the admissible states have zero 
norm. This is seen on determining those wave functions of ghost number zero that obey 
$\hat{\Omega}\psi(q,\lambda,\eta,\eta^{\times})=0$; they are obtained to be
\begin{equation}
\psi_{0;m}(q,\lambda,\eta,\eta^{\times})=\psi_{0;m}(q)\eta^{\times1}\cdots\eta^{\times 
m}\qquad\qquad
\psi_{m;0}(q,\lambda,\eta,\eta^{\times})=\psi_{0;m}(q)\eta^m\cdots\eta^1
\end{equation}
with $\hat{\varphi}_a\psi_{0;m}(q)=0$ and $\hat{\varphi}_a\psi_{0;m}(q)=0$. As follows 
from the Berezin rules, these wave functions indeed have vanishing norm. In such a 
situation it is usually argued in the literature that the norm is of the form 
$0\cdot\infty$ since the Grassmann integration gives $0$ and the integration over $q$ 
yields $\infty$ because one integrates over wave functions obeying 
$\hat{\varphi}_a\psi(q)=0$, and this results in an ill-defined expression. However, in view of what 
we have shown, this contradiction is void since the (super) trace is taken over the full 
extended state space $\mathcal{H}_{\mathrm{ext}}$; in addition, the trace is constructed 
by means of $\mathcal{H}_{\mathrm{ext}}$ and its dual $\mathcal{H}^{\ast}_{\mathrm{ext}}$ 
so that the (indefinite) inner product does not get involved at all. This is a fact of 
crucial importance, which is also confirmed by going through the proof of the Lefschetz 
formula.  Hence, we have a probabilistic interpretation because the states $\psi_{0;m}$ 
and $\psi_{m;0}$ are then paired in duality (if the relevant cohomology appears at a 
nonzero value $l$ one must invoke the duality $H^{+l}(\Omega)\cong H^{-l}(\Omega)$) so 
that
\begin{equation}
\langle\psi_{m;0}|\psi^{\prime}_{0;m}\rangle=\int\,dqd\eta 
d\eta^{\times}(\psi(q)\eta^m\cdots\eta^1)^{\ast}(\psi^{\prime}(q)\eta^{\times1}\cdots\eta
^{\times m})=\int\,dq\psi^{\ast}(q)\psi^{\prime}(q)
\end{equation}
where $\psi_{m;0}(q)=\psi(q)=\psi_{0;m}(q)$ (cf. also \cite{Mcmu 94}). Since the 
cancellation of the unphysical states in the (super) trace is automatic, there are also 
no normalization problems for the functional integral. This is the version of the no 
ghost theorem for a system with first class constraints.

What remains to resolve is the intriguing problem that the functional integral simultaneously takes care 
of all cohomology groups and not, as one would wish, of the zero 
cohomology only. One might argue that the operator $[\hat{\Omega},\hat{\phi}]$, or its 
exponentiated version entering the time-evolution operator, will be of special relevance. 
This operator has been introduced by Batalin and Marnelius (\cite{Bata 95}, see also 
\cite{Marn 91}) in the attempt to construct an inner product for constraints having a 
continuous spectrum. For the case at hand, it has recently been discussed by Rogers 
\cite{Roge 97}. In this latter work, the proposal is made that the operator 
$[\hat{\Omega},\hat{\phi}]$ could provide for the mechanism that only the zero cohomology 
survives in the partition function. This conjecture rests on the remarkable fact that, if 
the operator in question were invertible on all physical states, then there would be no 
cohomology at all; the proof is straightforward and amounts to showing that, under the 
above hypothesis, all physical states are also exact. The legitimate conclusion, as drawn 
by Rogers, is that the gauge-fixing fermion should be chosen such that the only states on 
which $[\hat{\Omega},\hat{\phi}]$ is not invertible are the elements of $H^0(\Omega)$. 
But in our case there is (up to the choice of $\chi$) no freedom in disposing of 
$\hat{\phi}$; beyond that, on the states $\psi_{m;0}$ and $\psi_{0;m}$ the action of 
$[\hat{\Omega},\hat{\phi}]$ is definitely nonzero since the operator 
$[\hat{\chi},\hat{\varphi}]$ is invertible; of course, the latter property only holds 
modulo Gribov obstructions \cite{Grib 78,Sing 78} (see also \cite{Baul 96}). Hence, for 
the case at hand (and, in particular, for Yang-Mills theory) this kind of approach does 
not work.

As we see it, there is no hope that the functional integral manages by itself that only 
the zero cohomology survives. Hence, it is forced upon us to introduce a thermodynamic 
potential $\gamma\in\mathbb{R}$ for the ghosts, and so we must consider the 
generalization
\begin{equation}
Z_{\chi}(t,\gamma)=\mathrm{Tr}_{\mathrm{B}}\mathrm{Str}_{\mathrm{GF}}\exp -
i(\hat{H}+[\hat{\Omega},\hat{\phi}]+i\gamma\hat{N})t
\end{equation}
which makes sense since the ghost number operator commutes with both $\hat{H}$ and 
$[\hat{\Omega},\hat{\phi}]$. Again, the partition function may be written as a path 
integral
\begin{equation}
Z_{\chi}(t^{\prime\prime}-t^{\prime},\gamma)=
\end{equation}
$$\int_{\mathrm{PBC}}\,d[p,q]d[\mu,\lambda]d[\zeta,\eta]
d[\zeta^{\times},\eta^{\times}]\,
\exp\,i\int_{t^{\prime}}^{t^{\prime\prime}}\,dt\left(p\dot{q}+\mu
\dot{\lambda}+i\zeta\dot{\eta}+i\zeta
\dot{\eta}-H\,-i\{\Omega,\phi\}-i\gamma N\right)$$
where the restriction to ghost number zero is obtained by expanding in terms of $\gamma$ 
and retaining the $\gamma$-independent term only. Hence, in the end, the thermodynamic  
potential for the ghosts is no longer visible; but it should tacitly be assumed as 
present.

\section{CONCLUDING REMARKS}

Finally, we want to comment on the subtleties that may arise in solving the Dirac 
condition for physical states, after the ghost degrees of freedom have been eliminated. 
As we shall see, topological properties in the large then get involved, which may give rise 
to anomalies of the quantum system.

We do this on the example of pure abelian Chern-Simons theory (see, e.g., \cite{Birm 91}) 
with the action
\begin{equation}\label{actioncs}
S=\frac{k}{4\pi}\int\,dt\int_{\Sigma} 
d^2x\,\varepsilon^{ij}\big(\dot{A}_i A_j + A_0 F_{ij}\big)
\end{equation}
where the two-dimensional domain $\Sigma$ is chosen to be either the whole plane or the 
torus. This is an action in Hamiltonian first order form so that the kinetic term 
determines the symplectic two-form, from which the canonical Poisson brackets can be read 
off to be (see also \cite{Fad 88,Jack 94})
\begin{equation}
\{A_i(x),A_j(y)\}=-\frac{2\pi}{k}\varepsilon_{ij}\delta(x-y).
\end{equation}
Since we want to quantize the system, a polarization \cite{Wood 92} must be chosen. There 
is no natural real polarization available, but the Levi-Civita tensor $\varepsilon^{ij}$ 
gives rise to a complex structure. Hence, we choose holomorphic quantization with 
\footnote{The conventions are $z=x^1+ix^2$, $\partial_z=\frac{1}{2}(\partial_1 - i
\partial_2)$ and $A_{\bar{z}}=\frac{1}{2}(A_1+iA_2)$.}
\begin{equation}
\hat{A}_{\bar{z}}=A_{\bar{z}}\qquad \hat{A}_z=\frac{\pi}{k}\frac{\delta}{\delta 
A_{\bar{z}}}
\end{equation}
and the Bargmann inner product for Schr\H odinger wave functionals 
$\psi[A_{\bar{z}}]$ then is 
\begin{equation}
<\psi_1|\psi_2>=\int \,d{[A_{\bar{z}},A_z]}\:\mathrm{exp}( -
\frac{k}{\pi}\int\,d^2x A_{\bar{z}}A_z)\:\overline{\psi_1[A_{\bar{z}}]}
\:\psi_2[A_{\bar{z}}].
\end{equation}
What the second term in \eqref{actioncs} tells us is that we have a system with the first 
class constraint 
\begin{equation}
\hat{C}=i\frac{k}{2\pi}\hat{F}_{12}=i\frac{k}{\pi}(\partial_{\bar{z}}\hat{A}_z
-\partial_z\hat{A}_{\bar{z}})
\end{equation}
where the time component $A_0$ of the gauge field serves as a Lagrange multiplier. Since,  
classically, the constraint $F_{12}=0$ only leaves gauge degrees of freedom, the system 
appears to be trivial, but quantum mechanically it is not if we quantize first and 
constrain afterwards. Furthermore, the Hamiltonian is identically zero; hence, there is 
also no evolution in time so that we face a purely cohomological problem. 

Physical wave functionals must obey the Dirac condition $\hat{C}(x)\psi[A_{\bar{z}}]=0$;  
but instead of trying to solve this by direct attack, we make a digression and look at 
the constraint as a symmetry transformation. By exponentiation, we obtain the operator
\begin{equation}\hat{U}[g]=\mathrm{exp}(-i\int\,d^2x\,\alpha\,\hat{C}) \end{equation}
with $g=\mathrm{exp}(-i\alpha)\in U(1)$, the action of which on Schr\"{o}dinger wave 
functionals is calculated to be
\begin{equation}\label{transflawswfunctionals}
\hat{U}[g]\psi[A_{\bar{z}}]=\mathrm{exp}(-iW[A_{\bar{z}};g])\;
\psi[g^{-1}A_{\bar{z}}]
\end{equation}
where $g A_{\bar{z}}=A_{\bar{z}}+\partial_{\bar{z}}\alpha$. Were it not for the 
exponential prefactor, this would be the standard behavior of the wave functional under 
time-independent gauge transformations. Instead, we have a projective transformation law 
with the Lie group 1-cochain (as opposed to the Lie algebra cochains having been 
encountered in the preceding sections)
\begin{equation}
W[A_{\bar{z}},g]=\frac{k}{\pi}\int d^2x 
A_{\bar{z}}\partial_z\alpha-
\frac{k}{2\pi}\int d^2x \partial_{\bar{z}}\alpha\partial_z\alpha
\end{equation}
which, potentially, is anomalous. We postpone the discussion of its properties and return 
to \eqref{transflawswfunctionals}; the integrated version of the Dirac condition then is
\begin{equation}
\hat{U}[g]\psi[A_{\bar{z}}]=\psi[A_{\bar{z}}].
\end{equation}
In the plane, this can be solved on passing to the complexification of $U(1)$; the result 
is 
\begin{equation}
\psi[A_{\bar{z}}]=\exp iW[A_{\bar{z}}]
\end{equation}
with
\begin{equation}
W[A_{\bar{z}}]=-i\frac{k}{2\pi}\int\,d^2xA_{\bar{z}}(x)\partial_zP(x,y)A_{\bar{w}}(y)
\end{equation}
where $P(x-y)=-4\partial_zG(x-y)$ denotes the Green's function of the operator 
$\partial_{\bar{z}}$, and $G(x-y)=-\frac{1}{4\pi}\mathrm{log}|\mu(z-w)|^2$ the standard 
propagator with $\mu$ an infrared cutoff. Despite the fact that the wave function solves 
the Dirac condition, nevertheless, it has finite norm with respect to the Bargmann inner 
product; thus, no normalization problems (cf. the remarks in the preceding section) 
arise.

Hence, the theory is exactly solvable but, as the transformation law 
\eqref{transflawswfunctionals} exhibits, the behaviour of wave functionals under gauge 
transformations is nonstandard. Whether it is truly anomalous or not, this depends on the 
1-cochain (for relevant background, see \cite{Kost 87,Jack 88}). So we need compute the 
coboundary 
\begin{equation}\label{boundary1cochain}
(\Delta W)[A_{\bar{z}};g,h]=W[g^{-1}A_{\bar{z}};h]+W[A_{\bar{z}};g]-
W[A_{\bar{z}};gh]=\frac{1}{2}[\hat{Q}[\alpha],\hat{Q}[\beta]]
\end{equation}
where $\hat{Q}[\alpha]=\int\,d^2x\,\alpha\,\hat{C}$. The right-hand side is the 
commutator of two abelian generators, and one expect this to vanish; if so, one can look 
at \eqref{boundary1cochain} as the integrated form of a Wess-Zumino consistency condition 
\cite{Wess 71}. 

In the plane, this is indeed correct. Moreover, the 1-cocycle is even exact since
\begin{equation}
W[A_{\bar{z}};g]=W[g^{-1}A_{\bar{z}}]-W[A_{\bar{z}}].
\end{equation}
Hence, we can pass to $ \psi^{\prime}[A_{\bar{z}}]=\exp\left(-
iW[A_{\bar{z}}]\right)\,\psi[A_{\bar{z}}]$ with conventional transformation law 
$\hat{U}^{\prime}[g]\psi^{\prime}[A_{\bar{z}}]=
\psi^{\prime}[g^{-1}A_{\bar{z}}]$ since the 1-coboundary disappears.

On the torus, however, the boundary of the 1-cochain is nonzero so that the Wess-Zumino 
consistency condition no longer holds. The reason is that, despite 
na\"\i ve expectation, the commutator on the right side of \eqref{boundary1cochain} 
need not vanish. This happens for large gauge transformations 
\begin{equation}g_m(x)\,=\,\exp\left(-i2\pi(m_1\frac{x_1}{L_1}+m_2\frac{x_2}{L_2})\right)
\end{equation}
where $m=(m_1,m_2)$ with $m_1$ and $m_2$ integer, which are not continuously connected to 
the identity. Here, the torus is considered as a rectangle $L_1\times L_2$ with opposite 
points identified; hence, boundary conditions become important. In particular, the 
correct choice of the generator of gauge transformations now is
\begin{equation}
\hat{Q}[\alpha]\,=\,\int\,d^2x\big(\partial_{\bar{z}}\alpha\,\hat{A}_z\,-
\,\partial_z\alpha\,\hat{A}_{\bar{z}}\big)
\end{equation}
which differs from the earlier form in decisive boundary terms; one then finds for the 
coboundary
\begin{equation}
(\Delta W)[A_{\bar{z}};g_m,g_n]=2\pi ik\,m\times n.
\end{equation}
The coupling constant generally takes values $k=r/s$ with $r$ and $s$ coprime integers;  
in particular, the case $r=1$ is of special relevance. This entails that, quantum 
mechanically, the product of two (classically commuting) large gauge transformations do 
not commute.

What this result shows is that pure abelian Chern-Simons in a finite geometry becomes 
truly anomalous since the abelian large gauge transformations, one begins with 
classically, no longer commute at the quantum level.

In concluding, let us mention this is not the end of the story. As has been shown 
elsewhere \cite{Gren 98,Gren 00}, the non-commutative behaviour of large gauge 
transformations leads to a quantum symmetry \cite{Char 94}. Hence, it appears that 
anomalies should also admit an interpretation in terms of quantum symmetries.

\section{APPENDIX}

In this appendix some formulae for Dirac states of ghost fermions over momentum space and 
Fourier transformation are collected (cf. also \cite{Salo 82}). We begin with the definition
\begin{equation}
|\zeta\rangle=\exp(\hat{\eta}\cdot\zeta)(\hat{\zeta}_m\cdots\hat{\zeta}_1)
|0\rangle\qquad\quad\quad
\langle\zeta|=\langle\bar{0}|\exp(\zeta
\cdot\hat{\eta})
\end{equation}
where now the Dirac ket has Grassmann degree $m$, and the dual bra has Grassmann degree 
zero. On these states, the momentum and coordinate operators act as follows
\begin{equation}
\hat{\zeta}_a|\zeta\rangle=\zeta_a|\zeta\rangle\quad\quad\hat{\eta}^a|\zeta\rangle=+\frac
{\partial_r}{\partial\zeta_a}|\zeta\rangle\qquad\qquad\langle\zeta|\hat{\zeta}_a=\langle
\zeta|\zeta_a\quad\quad\langle\zeta|\hat{\eta}^a=-
\frac{\partial_r}{\partial\zeta_a}\langle\zeta|.
\end{equation}
The normalization is
\begin{equation}
\langle\zeta|\zeta^{\prime}\rangle=(-1)^m\delta(\zeta-\zeta^{\prime})=(\zeta_m-
\zeta^{\prime}_m)\cdots(\zeta_1-\zeta^{\prime}_1)
\end{equation}
and the completeness relation takes the form
\begin{equation}
(-1)^m\int\,d^{m}\zeta\,|\zeta\rangle\,\langle\zeta|=1.
\end{equation}
The overlap with the configuration space basis turns out to be
\begin{equation}
\langle\zeta|\eta\rangle=\exp(\eta\cdot\zeta)\qquad\qquad\qquad\langle\eta|\zeta\rangle=
\exp(
\zeta\cdot\eta)
\end{equation}
and these bases are connected as follows
\begin{alignat}{2}
|\zeta\rangle& =(-1)^m\int\,d^{m}\eta \,\exp(-\zeta\cdot\eta)|\eta\rangle 
&\qquad\quad|\eta\rangle& =(-1)^m\int\,d^{m}\zeta\,\exp(-\eta\cdot\zeta)|\zeta\rangle\\
\langle\zeta|& =(-1)^m\int\,d^{m}\,\eta\exp(\zeta\cdot\eta)\langle\eta|&
\langle\eta|& =\int\,d^{m}\,\zeta\exp(\eta\cdot\zeta)\langle\zeta|.
\end{alignat}
We define $\psi(\zeta)$ through
\begin{equation}
|\psi\rangle=\int\,d^{m}\zeta\,|\zeta\rangle\,\psi(\zeta)
\end{equation}
and since $|\zeta\rangle$ is even, we have $\langle\zeta|\psi\rangle=(-1)^m\psi(\zeta)$ 
so that the Fourier transform and its inverse are given by
\begin{equation}
\psi(\zeta)=(-1)^m\int\,d^{m}\eta\,\exp(\zeta\cdot\eta)\psi(\eta)\qquad\qquad\qquad
\psi(\eta)=\int\,d^{m}\zeta\,\exp(\eta\cdot\zeta)\psi(\zeta)
\end{equation}
with the conventions $d\,^m\zeta=d\zeta^1\cdots d\zeta^m$ and $d\,^m\eta=d\eta^m\cdots 
d\eta^1$; this is the definition of the Fourier transform that is used in the main text.

\end{document}